\documentstyle[aps,prb,epsf,multicol]{revtex} 

\begin{document}

\title{The pair approach applied to kinetics in restricted geometries:
strengths and weaknesses of the method}

%
%

%
%
\author{Z. Konkoli\thanks{Corresponding author. %
        e-mail: zorank@fy.chalmers.se}$^{,1}$, 
        A. Karlsson$^2$, and O. Orwar$^3$}
\address{
    $^1${Department of Applied Physics, \\
         Chalmers University of Technology
         and G\"oteborg University, \\
         SE-412 96 G\"{o}teborg, Sweden} \\
    $^2${Department of Chemistry,\\
         G\"oteborg University, \\
         SE-412 96 G\"{o}teborg, Sweden} \\
    $^3${Department of Physical Chemistry \\
         and Microtechnology Centre at Chalmers,\\
 	 Chalmers University of Technology,\\
         SE-412 96 G\"{o}teborg, Sweden}
}

\date{\today}       
       
\maketitle       
       
\begin{abstract}      

In the rapidly emerging field of nanotechnology, as well as in biology
where chemical reaction phenomena take place in systems with
characteristic length scales ranging from micrometer to the nanometer
range, understanding of chemical kinetics in restricted geometries is
of increasing interest. In particular, there is a need to develop more
accurate theoretical methods.  We used many-particle-density-function
formalism (originally developed to study infinite systems) in its
simplest form (pair approach) to study two-species $A+B\rightarrow 0$
reaction-diffusion model in a finite volume.  For simplicity reasons,
it is assumed that geometry of the system is one-dimensional (1d) and
closed into the ring to avoid boundary effects.
The two types of initial conditions are studied with (i) equal initial
number of A and B particles $N_{0,A}=N_{0,B}$ and (ii) initial number
of particles is only equal in average $\langle N_{0,A} \rangle =
\langle N_{0,B} \rangle$.  In both cases it was assumed that in the
initial state the particles are well mixed.  It is found that particle
concentration decays exponentially for both types of initial
conditions.  In the case of the type (ii) initial condition, the
results of the pair-like analytical model agrees qualitatively with
computer experiment (Monte Carlo simulation), while less agreement was
obtained for the type (i) initial condition, and the reasons for such
behavior are discussed.
\end{abstract}   
     

\begin{multicols}{2}       
       
\narrowtext

\section{Introduction}

Classically, biochemical reaction kinetics is extrapolated from
measurements in dilute solutions and fitted into the cellular reaction
environment, and several flaws in this approach have been pointed
out.~\cite{IntRevCyto}  The main motivation of our work is to improve
the understanding of diffusion-controlled reactions in topologically
complex nanoscale environments represented in biological cells. In
this study attempt is made to develop better theoretical methods which
could describe diffusion-controlled reactions with boundaries. To
achieve this goal, a particular way of doing calculation, the
many-particle-density-function (MPDF)
approach,~\cite{PhysRev1,PhysRev2,kotkuz,kotkuz1} is modified to
account for presence of boundaries.

The theoretical findings of this study are relevant for experimental
work done in refs.~\onlinecite{Chiu1,Chiu2,Karl1}. Even if we focus on
biochemical reaction kinetics, the results should have an equal
bearing on nanotechnological applications such as
nanofluidics~\cite{Karl3} or molecular electronics~\cite{Park1}.  Both
are likely to be strongly dependent on reaction-diffusion behaviors of
molecules (nanofluidics) or electrons and holes (molecular
electronics) in restricted nanoscale geometries.

Most of the studies on diffusion-controlled reactions have been
performed for infinite systems without boundaries and a variety of
methods have been developed to do the analysis. The methods range from
mean field treatments towards more exact approaches which employ
quantum spin-chains~\cite{spinchains}, field theory~\cite{KJ1,KJ2}, or
MPDF formalism~\cite{PhysRev1,PhysRev2,kotkuz,kotkuz1}. The
references~\onlinecite{PhysRev2,ChemRev1} are an excellent review on
the subject. The opposite case when reactions take place in restricted
geometries with reactants confined into finite size, and eventually
squeezed into very small volumes, is less understood.  There is,
however, some pioneering work in this
area.~\cite{Khai1,Khai2,Tach1,Rama} In here, focus is on testing
performance of MPDF on diffusion-controlled reactions in finite
volumes.

The infinite diffusion-controlled systems posses quite remarkable
properties.  When dimensionality of the system $d$ is lower than some
critical dimension $d_c$ (e.g. for A+A reaction $d_c=2$~\cite{ben1},
and for $A+B$ $d_c=4$~\cite{ben2}) a new non-trivial sort of kinetics
sets in.  Taking $A+B\rightarrow 0$ as an example: Classical chemical
kinetic rate equation for this reaction, with initial densities equal
and homogeneous, is given by $\dot n(t)=-\lambda n(t)^2$ where dot
over symbol denotes time derivative.  For large time $t$ this equation
would predict density decay in the form of power law $n(t)\approx
{\cal A} t^{-\alpha}$. The amplitude of decay equals ${\cal
A}=1/\lambda$ and decay exponent is given by $\alpha=1$. In reality,
$\alpha=1$ holds only for a sufficiently high dimensionality of the
system when $d>4$, while for $d<4$ one has $\alpha=d/4$ and ${\cal
A}={\rm const} \sqrt{n_0} D^{-d/4}$, and ${\rm const}$ is just a
numerical factor. Please note that for $d<4$ decay amplitude ${\cal
A}$ does not depend on the reaction rate $\lambda$ and exhibits lack
of dependence on chemical details (universality).

The kinetics of the type described above is commonly referred to as
{\em anomalous} or {\em fluctuation-dominated}. The term ``anomalous''
points to the fact that mean field (or classical rate equations) fail
to describe such systems. The phrase ``fluctuation-dominated''
emphasizes importance of fluctuations in particle densities.  Once the
reaction creates a hole in the particle concentration, diffusion is
very slow in restoring the homogeneous particle density. This has to
do with recurrence of random walks. For $d\le 2$ probability that the
random walker will return to the same site after arbitrary number of
steps is equal to one. Random walkers tend to wander around their
initial position, and particles do not mix that well.  Rule of thumb
is that for lower dimensions the kinetics gets more anomalous.  Role
of dimensionality is well understood both for integer and
fractal (non-integer) like dimensions.~\cite{fractal}  On the other
hand, much less is known what happens when one shrinks the system
size, which is studied here.

To impact some progress in understanding reactions in restricted
geometry we analyze performance of MPDF approach and modify it to
account for finite reaction volume. To test such method of calculation
the A+B model is used as a study case. The A+B model is natural choice
for such a task. This model has been intensively studied for infinite
system
sizes.~\cite{kotkuz,ben2,burl1,ovch,thwil,burl2,gut,bram,oerd,mattis}
It was found that the A+B reaction has the remarkable property that
domains rich with A or B particles are formed as time goes on. Once
domains are formed the reactions happen only at domain boundaries
which leads to already mentioned decay exponent $\alpha=d/4$.  It is
interesting to study in what way the dynamics of the system (kinetics)
changes as one reduces the volume available to reaction, in particular
whether domain-like structure survives. For simplicity reasons, the
one-dimensional (1d) case is studied. In the calculation that follows
there is nothing special about 1d and present analysis can be easily
extended to the two- or three- dimensional cases.

The A+B model in restricted geometry has been studied before with the
assumption that one type of reactant is attached at the center of a
small volume, and it was further assumed that one type of particle is
in large excess.~\cite{Khai1,Khai2,Tach1} The more realistic problem
where all particles are allowed to move is much harder to solve, and
the goal of our study is to describe such a situation.  Also, in here,
the focus is on the case when the initial number of reactants is the
same, or roughly the same.  Naturally, the shape of the reaction
container might be important but this issue is not addressed at the
moment. To avoid boundaries completely, our 1d system will be closed
into the ring.

The paper is organized as follows. In section \ref{lattice} the model
is developed, {\em i.e.} detailed account is given of how particles
move and react. Lattice model is used due to its conceptual
simplicity. In section \ref{mpdf} equations of motion are derived
using MPDF formalism in its simplest form (pair approach).  In section
\ref{sec:solution} equations of motion are solved analytically and it
is shown how multi-exponential decay emerges.  The results of computer
experiments (Monte Carlo Simulations) are given in section
\ref{MCresults} followed by a comparison between theory and Monte
Carlo simulations in section \ref{comparison}.  We conclude by
analysis of strengths and weaknesses of the pair-approach applied to a
reactions in restricted geometry in section \ref{disc}.  In appendix
\ref{app:kt} details are given how to calculate effective reaction
rate $k(t)$ which determines density decay. In appendix \ref{app:MCS}
the algorithm used to do Monte Carlo simulations is discussed in
detail.

\section{The lattice model}
\label{lattice}

To test any theory one inevitably needs a model which serves as study
case. The model used here is defined as follows. The two species, A
and B, move on a 1d lattice performing random jumps with rates
(diffusion constants) $D_A$ and $D_B$ respectively. It is assumed that
$D_A=D_B$. Position of lattice sites is given by $x_i=i h$ with
$i=0,1,2,3,\cdots,M$ and $h$ denotes lattice spacing. Sometimes, $x$
and $y$ will be used instead of $x_i$.  Periodic boundary conditions
are assumed and sites $i=0$ and $i=M$ are defined to be
equivalent. There are M lattice sites in total and $L=M h$. By using
periodic boundary conditions it is possible to work with a system of
finite size and yet keep the spatial translational invariance. This
greatly facilitates the analytical treatment of the problem.

It is assumed that the reaction probability (per unit time) for
particle $A$ at $x$ and $B$ at $y$ is given by $\sigma(x-y)$. For 
$\sigma(x-y)$ simplest possible form is used
\begin{equation}
  \sigma(x-y)=\sigma_0 \theta(a-|x-y|).
  \label{sigma}
\end{equation}
where $\theta(x)=0$ for $x<0$ and $\theta(x)=1$ for $x\ge 0$.  In this
way two important aspects of chemical reactions are embedded, $a$
corresponds to the effective range of reaction and $\sigma_0$ is its
strength.  One could also say that each particle carries a ring of
radius $a/2$ and when two rings overlap the particles can react. In
this sense $a/2$ could be thought of as the size (radius) of
particles. For simplicity reasons it is assumed that the reaction
products do neither influence reactants, nor the A+B reaction. Also,
exclusion or steric effects are not taken into account, {\em i.e.}
particles are allowed to ``enter'' into each other (please see Fig. 1)
and react with same probability independently from which direction
they approach each other. 

\begin{figure}       
\epsfxsize=8cm       
\centerline{\epsfbox{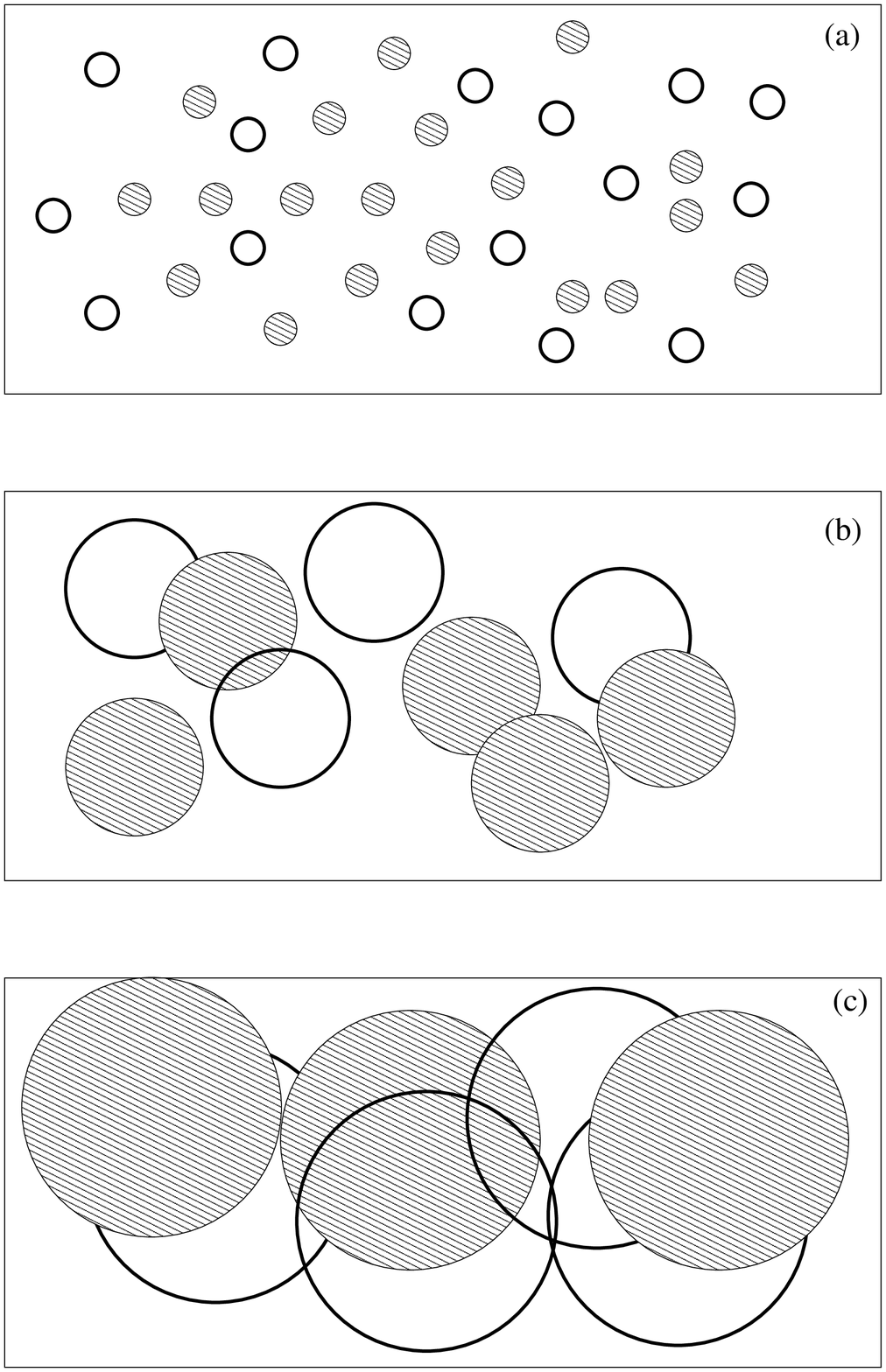}}       
\caption{Various situations which are simulated are shown. The three
figures schematically depict various types of initial conditions from
which simulation is started. (a) The upper most graph shows a
situation where particles react when nearest-neighbors only. The
reaction range is very short and particles come rarely in contact. (c)
The lowest figure shows a situation of dense packing with a large reaction
range. It corresponds to situation of high packing which occurs in a
cell environment. It is unrealistic that particles can penetrate into
each other but we consider this case nevertheless since it is simpler
to model. (b) The middle graph is midway between two extremes.}
\label{fig4}       
\end{figure}

The model has the useful property that if $a$ is thought of as the
size of reactants, then by varying $a$ several interesting situations
can be studied. For example, when $a$ is on the order of the system
size $L$, one can think of situations of extreme crowding.  On the
other hand when $a\ll L$ reactants appear as point-like objects. In
Fig. \ref{fig4} we offer a schematic way how to think about these
situations.  The model presented above is solved analytically and
numerically by a Monte Carlo simulation in the following sections.

\section{Equations of motion in pair approximation}
\label{mpdf}

To solve the A+B reaction-diffusion model in a restricted geometry we
use a many-particle-density-function formalism (MPDF), since it was
already used to describe asymptotics of the same reaction in an
infinite volume.~\cite{PhysRev1,PhysRev2,kotkuz,kotkuz1} We modify the
formalism and apply it to the case of a restricted geometry. In the
following the formulation presented in ref~\onlinecite{PhysRev1} will
be closely followed. On the way, the changes made to the original
formalism will be discussed.

The dynamics of the system, as defined in previous section, is
stochastic and governed by Master Equation which
describes time evolution of configurational probabilities of the
system $P(c,t)$,
\begin{equation}
  \dot P(c,t) = \sum_{c'} 
       \left[ 
          W_{c'\rightarrow c} P(c',t) - 
          W_{c\rightarrow c'} P(c,t) 
       \right]
  \label{ME}
\end{equation}
where $c$ is short notation for occupancy of lattice sites and
$W_{c\rightarrow c'}$ are transition probabilities which can easily be
deduced from the previous description of the model. Here and
throughout the paper dot over symbol denotes time derivative.

The quantities of interest are particle densities $\rho_A(x,t)$ and
$\rho_B(x,t)$ and they can be calculated from $P(c,t)$ (at least in
principle). Since system is closed into a ring translational
invariance holds and concentrations cease to be position dependent
which leads to $\rho_A(x,t)=n_A(t)$ and $\rho_B(x,t)=n_B(t)$.
Following recipe in ref.~\onlinecite{PhysRev1} gives following equations for
$n_A$ and $n_B$,
\begin{eqnarray}
  & & \dot n_A(t) = - n_A(t) n_B(t) \int_{-L/2}^{L/2} dx \sigma(x) Y(x,t) 
      \label{dotnA} \\
  & & \dot n_B(t) = - n_A(t) n_B(t) \int_{-L/2}^{L/2} dx \sigma(x) Y(x,t) 
      \label{dotnB}
\end{eqnarray}
The $Y(x,t)$ denotes correlation function for AB pairs. Absence of
correlations is signaled by $Y(x,t)=1$. Please note that in this work
system size is finite which enters through finite integration domain
in the integrals above (it might appear as minor technical detail but
this fact is very important).  Also, it is assumed that reversal
symmetry holds, i.e. $Y(x,t)=Y(-x,t)$.

Again, following ref.~\onlinecite{PhysRev1} one can derive equation for
$Y(x,t)$ which is given by
\begin{eqnarray}
   & & \dot Y(x,t) =  
      (D_A+D_B) Y''(x,t) - \sigma(x) Y(x,t) \nonumber \\
     & &  - n_B Y(x,t) \int_{-L/2}^{L/2} dy \sigma(y) Y(y,t) 
          \left[ X_B(x-y,t) - 1 \right] \nonumber \\
     & &  - n_A Y(x,t) \int_{-L/2}^{L/2} dy \sigma(y) Y(y,t) 
          \left[ X_A(x-y,t) - 1 \right]
     \label{dotY}
\end{eqnarray}
where prime denotes spatial derivative, and $X_A(x,t)$ and $X_B(x,t)$
correlation functions for AA and BB pairs respectively. The $X_A(x,t)$
and $X_B(x,t)$ obey similar equations which are not given here to
save the space.

The equations (\ref{dotnA})-(\ref{dotY}) are derived under assumption
of Kirkwood superposition approximation, which is a technical way of
saying that dynamics is governed by pair effects.  Naturally,
assumption of the dominance of pairs effects is an approximation. It
might or might not work, and the goal of present study is to test
this. In the following, to make analytic treatment possible, equations
will be simplified further by setting $X_A(x,t)$ and $X_B(x,t)$ equal
to one. This amounts to ignoring correlations among AA and BB
pairs. In ref.~\onlinecite{PhysRev1} it was shown that (for infinite
reaction volume) such approximation is too severe and does not lead to
correct decay exponent $\alpha=d/4$ (it gives
$\alpha=d/2$). Nevertheless, in here we consider such
simplification. The validity of such an approximation, together with
the fact that we are using Kirkwood approximation, is tested via
computer experiment later on.

The form of boundary conditions for $Y(x,t)$ differs from the one used
in ref.~\onlinecite{PhysRev1}. In the case of infinite system one takes
\begin{equation}
  Y(x,t)\rightarrow 1 \ , \ \ x\rightarrow\infty
  \label{XYinf}
\end{equation}
while for finite system with periodic boundary conditions another form
has to be used
\begin{equation}
  Y(x+L,t) = Y(x,t) \label{XYfin}
\end{equation}
It will be shown later that the change from (\ref{XYinf}) to
(\ref{XYfin}) leads to a qualitative change from power law to (multi)
exponential behavior for correlation dynamics.  The rest of the
boundary conditions are standard, and are taken as in the case of an
infinite system size,
\begin{eqnarray}
  & & n_A(0) = n_0  \label{nA0} \\
  & & n_B(0) = n_0  \label{nB0} \\
  & & Y(x,0) = 1 \label{XY0}
\end{eqnarray}
Also, taking $L\rightarrow\infty$ should reproduce findings of
ref.~\onlinecite{PhysRev1}, within set of approximations employed
here.

\section{Emergence of multi-exponential density decay}
\label{sec:solution}

With assumptions $X_A=X_B=1$ Eq.~(\ref{dotY}) reduces to
\begin{equation}
  \dot Y(x,t) =  
    (D_A+D_B) Y''(x,t) 
      - \sigma(x) Y(x,t) \label{dotYsm}
\end{equation} 
Eq.~(\ref{dotYsm}) is solved by using a Laplace transform as shown in
the appendix \ref{app:kt}. To simplify the algebra, it is assumed that
$\sigma_0$ is arbitrary large. The exponential behavior emerges due to
the fact that the spectrum of Eq.~(\ref{dotYsm}) is discrete due to
particular nature of the boundary conditions. The final expression for
$k(t)$ reads
\begin{equation}
  k(t) = k_{reg}(t) + 2 a \delta(t)
  \label{kt}
\end{equation}
and details of calculation are given in the appendix \ref{app:kt}.
The regular part of $k(t)$ is given by
\begin{equation}
  k_{reg}(t) = \frac{8D}{L-2a}\sum_{m=1}^{\infty} e^{-\kappa_mDt} 
  \label{kreg}
\end{equation}
and $\kappa_m$ are constants of multi-exponential decay (eigenvalues),
\begin{equation}
  \kappa_m = \pi^2 \left( \frac{2m-1}{L-2a} \right)^2 
  \label{kappam}
\end{equation}
The $\delta$-function term in (\ref{kt}) arises from the second term
on the right hand side of Eq.~(\ref{ks}) when
$\sigma_0\rightarrow\infty$ (please see the appendix \ref{app:kt}).

Once $k(t)$ is available one can calculate $n(t)$ as
\begin{equation}
  n(t) = \frac{n_0}{1+I(t)n_0+2 a n_0}
  \label{nt}
\end{equation}
where $I(t)=\int_0^t dt' k_{reg}(t')$ and 
\begin{equation}
  I(t) = \frac{8}{L-2a}
         \sum_{m=1}^\infty\frac{1}{\kappa_m}
         \left( 1 - e^{ -\kappa_m D t } \right)
  \label{It}
\end{equation}
The $2an_0$ term in the denominator of (\ref{nt}) comes from the
$\delta(t)$ term in Eq.~(\ref{kt}). It describes the immediate
annihilation of particles which are within reaction range. When
$\sigma_0\rightarrow\infty$, this happens instantaneously. Thus there
is a sudden jump in particle concentration. For finite $\sigma_0$
this jump becomes a smooth transition (exponential decay with decay
exponent proportional to large number $\sigma_0$).

The question is whether one can obtain results for an infinite system
from the Eq.~(\ref{nt}) above. This can be done using Poisson
resummation formula when $Dt/(\frac{L}{2}-a)^2\ll 1$. The Poisson
resummation procedure gives $I(t)\sim t^{1/2}$ which results in the
wrong exponent for the density decay; $n(t)\sim t^{-1/2}$ instead of
correct $n(t)\sim t^{-1/4}$.  Thus we just reconfirm the well known
fact that for infinite systems, the pair approach predicts too fast
decay of particles.  However, for finite systems, the situation is not
that clear, it appears to depend on the type of initial conditions the
real system is subject to.

In here we consider two types of initial conditions. (i) When
initially there is an equal number of A and B particles, one has
$n\rightarrow 0$ as $t\rightarrow\infty$; and at the end all particles
have to annihilate. (ii) One can look at an ensemble of similar
systems with equal number of A and B particles at $t=0$ on average,
$\langle N_{0,A}\rangle = \langle N_{0,B} \rangle$. In such a case,
one has different asymptotics, $\langle N(t) \rangle\rightarrow
N(\infty)$ as $t\rightarrow\infty$.

Theoretical prediction is that, as time goes to infinity, the particle
density exponentially approaches the value $n_{th}(\infty)$;
\begin{equation}
  n_{th}(\infty)=\frac{n_0}{1+n_0L}
  \label{nthassy}
\end{equation}
The value for $n_{th}(\infty)$ above can be obtained by sending
$t\rightarrow\infty$ in Eqs.~(\ref{nt}) and (\ref{It}).  From
(\ref{nthassy}) one sees that asymptotically number of particles is
given by 
\begin{equation}
  N_{th}(\infty)=N_0/(1+N_0)
  \label{Nthassy}
\end{equation}
where $N_{th}(\infty)=Ln_{th}(\infty)$. Please note that
$N_{th}(\infty)$ never approaches zero and settles at a number between
zero and one.  In the case of type (i) initial conditions, all
particles annihilate and $N(\infty)=0$. This is clearly in
contradiction with Eq.~(\ref{nthassy}). However, situation is not that
hopeless, as will be discussed later.  For the type (ii) initial
condition, for each member in ensemble there is a chance that more
than one particle will be left, since one start dynamics with (random)
excess of A or B particles at $t=0$. Thus, in average, $\langle
N(\infty) \rangle$ will be larger than 0. Clearly, pair approach has
more chance to describe this situation correctly.

In summary, we find an exponential decay in the long time limit which
is a pure artifact of the finites of the system. There is a clear
indication that the quality of prediction depends on the type of
initial conditions used in experiment. Also, the approximations made
in deriving Eq.~(\ref{nt}) are rather severe and in order to check the
applicability of such a pair-approach Monte Carlo simulation is used.

\section{Results of Monte Carlo simulations of A+B reaction 
in restricted geometry}
\label{MCresults}

\begin{figure}       
\epsfxsize=8cm       
\centerline{\epsfbox{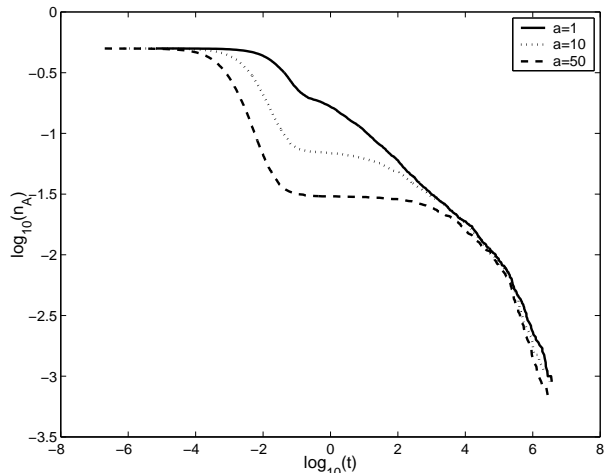}}       
\caption{Result of Monte Carlo simulations in 1d for type (i) initial
condition. A very large system is simulated on a lattice with $L=10^4$
sites. Also, the initial number of particles $N_{0,A}=N_{0,B}=5000$ is
very large. Simulation starts from the largest possible density
$n_{tot}(0)=1$~particle/site. A and B particles have the same diffusion
constant $D_A=D_B=1 s^{-1}$. Asymptotically, the number of particles
approaches zero. There are three distinct regimes present; (a) of the
mean field decay ($-\infty<log_{10}(t)<-2$), (b) plateau where
particle concentration does not change much ($-2<log_{10}(t)<2$), (c)
power law decay ($2<log_{10}(t)<5$), and (d) exponential decay at the
end $5<log_{10}(t)<\infty$. The indicated ranges are given roughly
just to guide the eye. They also depend on which $a$ is used in
simulation.}
\label{fig1}       
\end{figure}

\begin{figure}       
\epsfxsize=8cm       
\centerline{\epsfbox{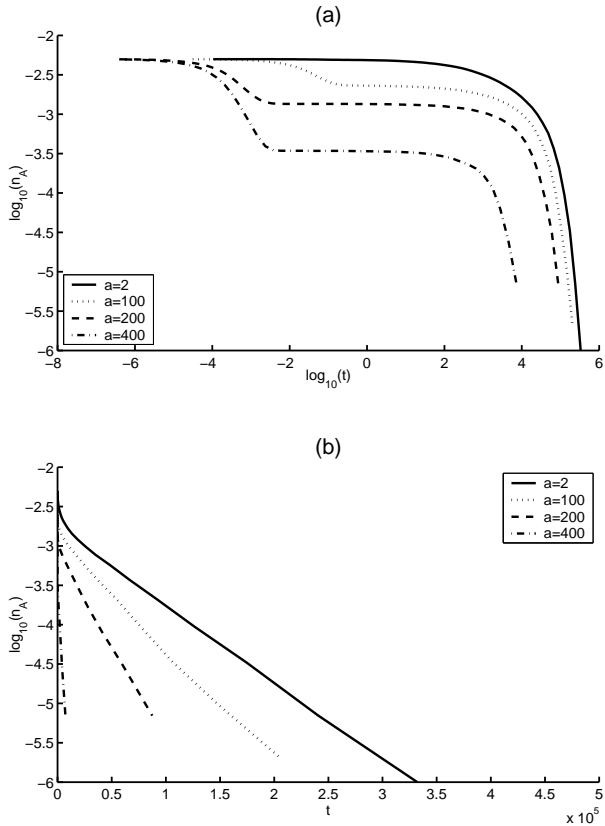}}       
\caption{Study of the exponential regime where small number of
particles is present on the lattice for type (i) initial condition
with $N_{0,A}=N_{0,B}=5$. The number of lattice sites is $L=1000$. All
other parameters are same as in figure 1. Each curve is obtained as
average over 1000-3000 runs. Asymptotically, the number of particles
approaches zero. Panel (a) shows log-n versus log-t plot to trace down
power law decay (should appear as a straight line). There is no power
law decay. Also, small and large $t$ region are resolved better. Panel
(b) shows log-n versus t plot to indicate exponential decay
(corresponds to straight lines). The particle density vanishes
exponentially $n\sim{\rm exp}(-\kappa t)$ where $\kappa$ depends on $a$
since the slopes for all curves are different. There is a value
$a=L/2$ when $\kappa$ becomes infinite (particles can not escape each
other).}
\label{fig2}       
\end{figure}

\begin{figure}       
\epsfxsize=8cm       
\centerline{\epsfbox{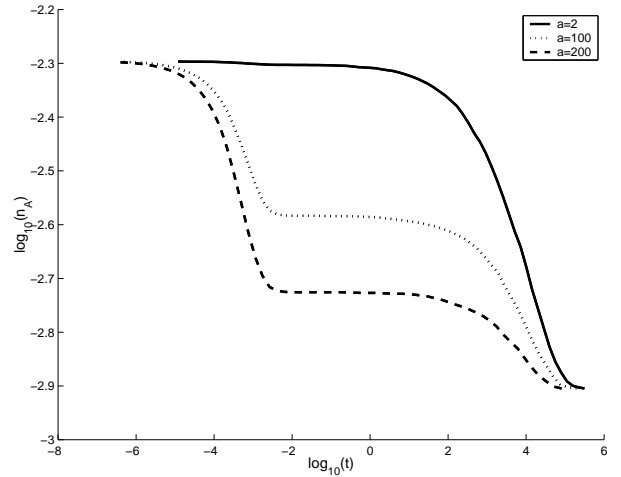}}       
\caption{Simulation for type (ii) initial condition. All parameters as
in the figure 2. The only difference from figure 2 is in the initial
condition. $N_{0,A}$ and $N_{0,B}$ vary randomly with constraint that
$N_{0,A}+N_{0,B}$ is fixed and equals $10$. Asymptotically, number of
particles does not approach zero.  }
\label{fig3}       
\end{figure}

Figures \ref{fig1}, \ref{fig2}, and \ref{fig3} summarize the results
of the Monte Carlo simulations in $d=1$. The Monte Carlo algorithm is
described in detail in appendix \ref{app:MCS}. Figure \ref{fig1} shows
a simulation for a system with a large initial number of particles
with a varying reaction range from a nearest-neighbor interaction with
$a/L=0.0001$ towards a longer range with $a/L=0.02$.  Figure
\ref{fig2} shows the case when there are initially very few (exactly
10=5A+5B) particles present in the reaction volume, also with varying
reaction ranges from $a/L=0.001$ to $a/L=0.2$. Thus figures \ref{fig1}
and \ref{fig2} give simulation results for type (i) initial
condition. Figure \ref{fig3} deals with type (ii) initial condition,
when the initial number of particles in an ensemble varies with the
constrain that the sum of A and B particles equals $10$. (For example,
one run could be done with 7A and 3B, the other run with 5A and 5B,
and a third run with 4A and 6B particles, etc.) Figure \ref{fig4} is a
sketch of how to think of various situations when $a$ changes from
small to large values.

From figure \ref{fig1} it can be seen that in the case of the nearest
neighbor reaction range ($a=1$) four distinct regimes appear and the
log-log plot is used to reveal them; (a) mean field decay, (b) the
plateau region (c) power law decay and (d) exponential decay at the
end.  These regimes disappear as the reaction range is increased, and
eventually, for very large $a$, one only has the exponential regime.

The mean field regime corresponds to annihilation of particles with
all reactants being well mixed.  This leads to depletion of lattice to
concentration of the order $n\sim 1/a$, thus one particle per reaction
range. Then diffusion starts to operate and mixes particles.  What is
interesting is that for very large values of $a$, the plateau region
starts earlier and lasts longer.  Apparently, it takes some time
before the particles find each other by diffusion and start reacting
again.

The power law decay starts after the plateau region. There is
universality in the power-law regime since all curves with different
values of $a$ merge into one.  This is somewhat surprising since a
larger $a$ should mean faster annihilation, which indeed happens in
the mean field regime, but yet in the power law stage all curves share
the same power-law behavior. We speculate that this has to do with
self organization and build up of correlation.

The exponential regime is entered after the power law regime, when the
number of particles in the system becomes small. With the present
computer hardware it was not possible to resolve this exponential
regime better. This is indeed done in figure \ref{fig2} with a smaller
lattice size and lower particle number.

To illuminate this exponential decay at later stages of annihilation,
we performed simulations with a smaller number of particles (10=5A+5B)
on a smaller lattice with $10^3$ sites. Thus we used type (i) initial
condition. To obtain each curve we followed 1000-3000 realizations of
dynamics and averaged over such an ensemble. The result is shown in
figure \ref{fig2}. The upper figure is in log-$t$ scale to resolve the
small and large $t$ region, respectively. The lower figure is in
normal-$t$ scale and we use it to detect exponential decay (where a
straight line indicates exponential decay).

The crossover from mean-field to plateau-like dynamics can be seen in
the upper graph where all curves drop down to a plateau value which is
$a$-dependent. The theoretical prediction for this plateau is
$n(0^+)=n_0/(1+2 a n_0 )$.  The initial drop in concentration is large
for large $a$-values.  In the upper figure, it is hard to say when the plateau
behavior turns into exponential decay.

The lower graph shows that decay is indeed exponential since density
curves at late times are straight lines in the log$_{10}(n)$-$t$
plot. Thus at the late times $n\sim{\rm exp}(-\kappa_1 t)$. Also, the
decay constant $\kappa_1$ is $a$-dependent since slopes are different
for various values of $a$, and $\kappa_1$ becomes larger with larger
$a$. Also, it appears that there is an upper limit for $a$ at which
decay becomes infinitely fast. Naturally, this happens when $a=L/2$
since none of the particles can escape from each other. The
qualitative dependence of $\kappa_1$ on $a$ just discussed is in
agreement with theoretical prediction in Eq.~(\ref{kappam}) with
$m=1$.


Figure \ref{fig3} is obtained in a similar way as figure
\ref{fig1}. The only difference is that figure \ref{fig3} deals with
the type (ii) initial condition. For the particular run, when
$N_{0,A}\ne N_{0,B}$, the final number of particles in the system is
not zero. For example, when starting from 7A and 3B particles, the
system will end up in the state of 4A particles. This comes from that
fact that the A+B reaction conserves the particle difference
$N_A(t)-N_B(t)=const$.  Curves for different values of $a$ saturate at
one single value which is, independent of $a$. Clearly, the value of
the plateau is solely controlled by the excess of particles at $t=0$,
and can be calculated from theory if needed, but result of Monte Carlo
simulation is equally informative.

\section{Comparison of computer experiment and theory}
\label{comparison}

\begin{figure}       
\epsfxsize=8cm       
\centerline{\epsfbox{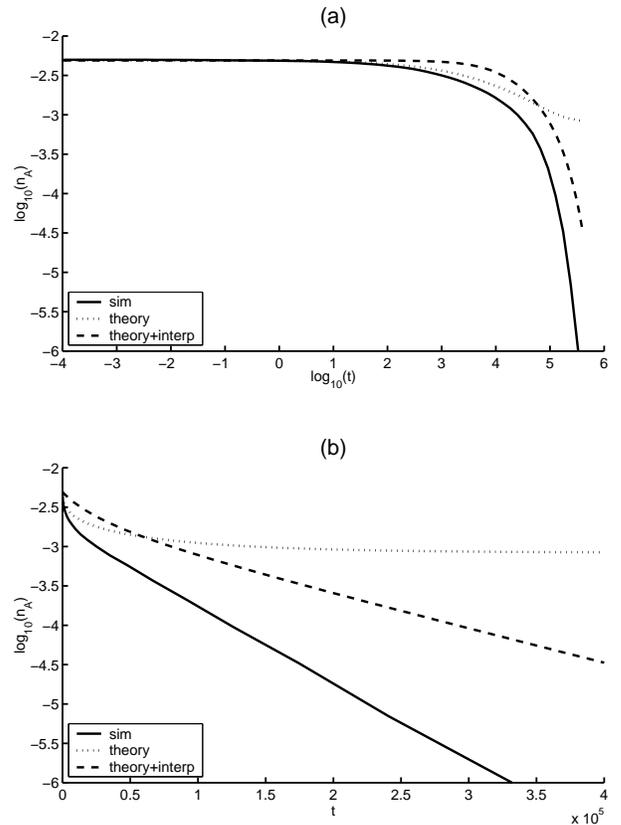}}       
\caption{Comparison of theory and experiment (Monte Carlo simulation)
for type (i) initial condition. Simulation data are taken from figure
2. Panel (a): theory (dotted line, Eq.~\ref{nt}) predicts
$\lim_{t\rightarrow\infty}n(t)\ne 0$ while in reality
$n(t=\infty)=0$. Reasons for this discrepancy are given in the
text. Panel (b): by using interpolation formula (\ref{nint}) one
obtains dashed curve. Agreement with simulation gets better.}
\label{fig5}       
\end{figure}

Figure \ref{fig5} shows a comparison of the analytical treatment with
computer experiment (simulation parameters as in figure
\ref{fig2}). It can be seen that the pair (Smoluchowskii) approach
does not predict that the number of particles in the system should
approach zero. The reasons for this are discussed later but have to do
with the fact that we are looking at highly symmetric situation with
equal number of A and B particles all the time.  To enforce such zero
asymptotics by hand we use interpolation formula
\begin{equation}
  n_{int}(t) = n(t) - n(\infty)(1-e^{-\kappa_1 t})
  \label{nint}
\end{equation}
where $\kappa_1$ is the first dominant large time exponent in expression
for $k(t)$ (see Eqs.~\ref{nt} and \ref{It}). It can be easily seen
that the equation above holds exactly for $t=0$ and $t=\infty$. If
Eq.~(\ref{nint}) is used instead of (\ref{nt}) the agreement with
simulation improves in the sense that the decay is exponential and
the qualitatively theoretical exponent is roughly the same as the one
obtained from simulations. More work is in progress to develop improved
interpolation formulas.

\begin{figure}       
\epsfxsize=8cm       
\centerline{\epsfbox{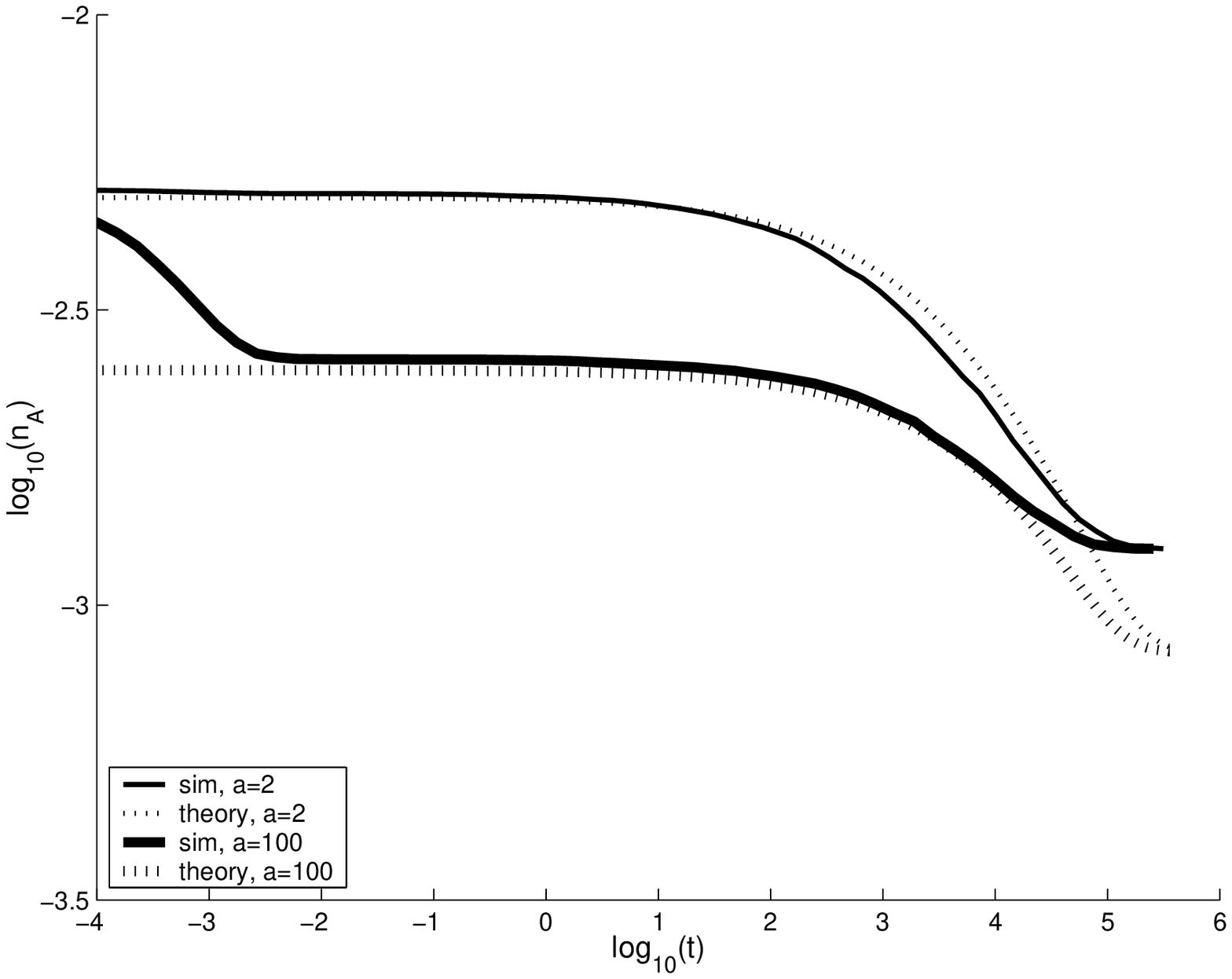}}       
\caption{Comparison of theory and experiment (Monte Carlo simulation)
for type (ii) initial condition. Both theory and simulation give
$n(\infty)\ne 0$. Theory (dotted line, calculated with Eq.~\ref{nt})
predicts faster annihilation of particles.}
\label{fig6}       
\end{figure}

Figure \ref{fig6} deals with the same type of comparison, but with a
simulation setup as in figure \ref{fig3} when the initial number of
particles is not fixed, just the total number (type (ii) initial
condition). One can see that the agreement between theory and
simulation is much better. Clearly, Smoluchowskii theory deals better
with the type (ii) initial condition represented by the figure
\ref{fig3} than by the type (i) represented by the figure
\ref{fig2}. Also, in figure \ref{fig6}, one can see that theory
predicts too fast particle annihilation. This is no surprise since
this is what one would expect form such pair approach which does not
take into account formation of domains. (In that respect there is
similarity with infinite systems, but only for the type (ii) initial
condition).

\section{Discussion}
\label{disc}

The goal of present work was to impact some understanding of 
diffusion-controlled reactions in restricted geometries, with aim to
describe some aspects of chemical reactions in biological cells. Two
issues have been dealt with:

(1) The particular way of doing calculation was tested, the MPDF
formalism developed by Kuzovkov and Kotomin (see ref.~\onlinecite{PhysRev1}
for details). To be able to solve equations analytically the hierarchy
of many-particle-densities was truncated at the level of
three-particle-density using shortened Kirkwood superposition
approximation. This approximation amounts to assuming that pair
effects dominate correlations, and present calculation can be viewed
as a variant of pair approach.

(2) A two species reaction-diffusion model $A+B\rightarrow 0$ in a
restricted geometry was taken as a study case.  Two types of initial
conditions were considered, type (i) where the initial number of A and
B particles is strictly equal, and type (ii) initial conditions where
the initial number of particles is equal only approximatively.  

Thus the paper is best viewed as a method paper since the main goal is
to test the strengths and weaknesses of the pair approach.  To test
quality of approximations involved all results have been compared with
the results of computer experiment (Monte Carlo simulation).

From a theoretical point of view, it seems that the pair method, being
widely used in calculation of bulk properties, works with mixed
success for the restricted reaction diffusion systems, at least the
one studied in here. The agreement between theoretical calculation and
computer experiment is qualitative in the case of type (ii) initial
conditions.  In the case of type (i) initial conditions there is less
agreement, however, situation is not that hopeless.

In the case of type (i) initial conditions pair approach makes error
of the order of one particle (please see Eq.~\ref{Nthassy}), since it
predicts that, in the average, after very long time, there will be
between zero and one particle in the system (though all particles
should vanish). When initial number of particles is
relatively large, the pair approach can describe evolution of system
for rather long time, before the regime is reached where only one
particle is left in the reaction volume.  However, in the case of
small initial number of particles, there is no such time interval, and
the mismatch between pair approach and simulation has to be addressed
more seriously.

The weakness of pair approach in dealing with type (i) initial
condition rests on the fact that the truncated set of equations for
many-particle-densities do not recognize any effects which go beyond
pair correlations. For example, in the present case, all information
related to the fact that initially there were 5A and 5B particles in
the system, and that all particles have to vanish eventually, is
missing.  Work is in progress to pass such type of information from
higher order particle-density-functions to lower order ones.  For
example, we already have better interpolation scheme than the one
given in Eq.~(\ref{nint}) for the case when there are initially three
particles in the system, but we are trying to understand how to extend
such analysis to higher numbers.

Interestingly enough, it seems that, contrary to the systems with
infinite sizes, setting $X_A(x,t)=X_B(x,t)=1$ is reasonable
approximation for a finite system, but we have to perform more tests.
This could have to do with the fact that if the system is too small,
there will be no time to develop clusters of A and B particles, and
setting $X_A(x,t)=1$ and $X_B(x,t)=1$ might turn to be a good
approximation after all. Thus, one does not have to turn to more
complicated methods of calculation if qualitative results are needed.
Nevertheless, it is highly desirable to see what happens as one
includes correlations among AA and BB pairs.

To be able to solve equations analytically, we had to simplify MPDF
considerably down to the level of a pair like approach and various
calculation schemes contain pair approach as possible
approximation. Perhaps the most common form of pair approach is the
one suggested by Smoluchowskii (see e.g. ref.~\onlinecite{PhysRev2} for
interesting review). The Smoluchowskii approach boils down to solution
of Poisson equation with different boundary conditions. (Many readers
will be familiar with this in the context of heat transfer or quantum
mechanical problems.)

However, one has to keep in mind that pair approach is an
approximation, and it has to be tested to see whether it works. (For
example, the pair approach can not describe A+B reaction-diffusion
model when system size is very large. This has been discussed in
ref.~\onlinecite{kotkuz}.) The advantage of pair approach, in the form used
here, is that it is possible to go beyond it in a systematic way. 

Also, from the particular way we have approached pair problem, one can
see that the difficulties associated with Eq.~(\ref{Nthassy}) are
likely to be much deeper than just the fact that we are using pair
approach. (This problem clearly vanishes when system size is infinite,
as $n_{th}(\infty)$ goes to zero.) Any scheme which focuses on low
rank particle-density-function will suffer in a similar way, in the
case of the highly symmetric, e.g. type (i), initial condition. One
really has to find a way how to incorporate information of higher
order correlation functions into lower order ones, without calculating
higher order correlations functions explicitly. This is a pressing
issue.

A few words about the model used. The goal of present work is to
develop calculation method rather than to describe specific chemical
system. We used the model which we could solve, within reasonable
level of approximation.  Nevertheless, the question whether present
model has any relevance for real biological and chemical systems needs
to be addressed.

The reaction diffusion model studied here appears to be too simple,
for two reasons. First, it does not account for chemical details which
enter only through two parameters, diffusion constant $D$ and reaction
rate $\lambda$. For example, the exclusion effects and steric effects
are not contained in it. Also, the influence of product molecule is
completely ignored. Second, 1d character of the model might be too
restrictive.

Despite its simplicity, the model used here contains basic
characteristics of diffusion-controlled reactions (reaction times
$t_R$ are much smaller than the corresponding diffusion times $t_D$);
particles are moving on the on the lattice and react when within
reaction range with no memory of initially velocity.  Previous
research reviewed in refs.~\onlinecite{PhysRev1,PhysRev2} has shown
that diffusion-controlled models, similar to the one used here, can be
used to describe real chemical reactions. In particular, the A+B model
have been used to study two reactions in capillary tube quite
successfully, bromine + cyclohexene $\rightarrow$ adduct, and
Cu$^{2+}$ + disodium ethyl bis(5-tertrazolylazo)acetate trihydrate
$\rightarrow$ 1:1 complex in water.~\cite{Kopelman}

Why can one be sloppy and ignore chemical details to some extent? The
reason for this is universality. Most often, predictions of the
reaction-diffusion models (on lattice) are insensitive to the details
of the chemistry involved. (For example, the decay amplitude for A+B
reaction does not depend on the reaction rate $\lambda$.) This
statement is valid provided one deals with very large system sizes.
Naturally, this view is not the only one. There are other ways to
approach diffusion-controlled reactions. For more chemical or
biological approach to diffusion-controlled reactions see
refs.~\onlinecite{ChemRev1,ChemRev2} and~\onlinecite{BiolRev}
respectively.  

The simplicity of the model is not necessarily such a big handicap,
until one reaches extremely small sizes.  For small system sizes,
density decay will start depending on details.  However, there is a
large window in system sizes between extremely large and extremely
small where such kind of universality could survive.  In here we push
the model over its borders by studying situation of extreme crowding
and not accounting for exclusion effects.

Furthermore, we would like to notify the reader that we do refer to
the model here as a ``toy'' model, it relatively simple to formulate
it. However, this does not mean that the model is easy to solve, quite
the contrary. In the case of infinite system size it has taken a lot
of research effort to clarify that the decay exponent indeed is
$d/4$. This issue was finally settled in ref.~\onlinecite{bram} which
provides a strict mathematical proof.

The model has a potential to describe experiments in
refs.~\onlinecite{Chiu1,Chiu2,Karl1} where, for example, the average
diameter of the reaction container (liposome) is $L\sim 1-25 \mu
m$. The reactants A (enzyme) and B (substrate) are of the size of
$a_{E},a_{S}\sim 1 nm$ and the typical number of reactants inserted is
on the order of $N\sim 1000$. Thus, $a\ll L$ holds to a very good
approximation. In these experiments reactants appear as point-like
objects and there is no need to give structure to reactants. Also,
problems associated with Eq.~(\ref{Nthassy}) will likely not to case
any damage due to large number of particles at $t=0$.

From the data above, the concentration of particles $n_{A,B}\sim
N/L^3$, is easily estimated to be $n\sim 1 \mu m^{-3}$, and the
typical distance between particles is $d_{AB}\approx 1/c^{1/3} \approx
L/N^{1/3} \sim 0.1 \mu m$. Thus, in average, $a_{S,E}\ll d_{AB}$, and
particles are very well separated. Therefore, it is reasonable to
assume that pair effects are the dominant ones. This in turn
simplifies the theoretical description considerably. Clearly, there
are other scales in the problem and the criteria on applicability of
the pair approach are more subtle in reality.~\cite{PhysRev2,kotkuz}

To summarize, it would be extremely interesting to have a general
method of calculation which could describe diffusion-controlled
reactions in finite volumes, perhaps something on the level of the
pair approach.  Pair approach is attractive since inclusion of
chemical details such as exclusion or steric effects is possible (see
e.g. work in ref.~\onlinecite{steric}) which certainly opens a new
route towards more quantitative results. However, pair approach is an
approximation, and before burdening pair approach with increasing
amount of chemical details, one has to test its ultimate
reach. Present work is an attempt in this direction.

\appendix

\section{Derivation of the reaction rate $\lowercase{k(t)}$}
\label{app:kt}

With initial condition $Y(x,0)=1$, the Laplace transform of
(\ref{dotYsm}) becomes
\begin{equation}
  s Y(x,s) - 1 = (D_A+D_B) Y''(x,s) - \sigma(x)Y(x,s) 
  \label{dotYsm1d}
\end{equation}
The correlation functions are symmetric around origin,
i.e. $Y(x,t)=Y(-x,t)$. They are also periodic in $L$. This implies that
it is sufficient to focus on positive x axis and impose boundary
conditions $\frac{\partial}{\partial x}Y(x,s) =0$ for $x=0$ and
$x=L/2$. Equation (\ref{dotYsm1d}) is an ordinary second order
differential equation, which is easily solved by solving it in regions
$0<x<a$ and $a<x<L/2$ separately and then matching solutions at the
end. After some algebra one obtains
\begin{eqnarray}
  & & Y_1(x,s)  = \frac{1}{D}
                 \left( \frac{1}{\nu}-\frac{1}{\mu} \right) \times \nonumber \\
  & & \ \ \ \times  \frac{
                {\rm ch}(x\sqrt{\mu})
              }{
                {\rm ch}(a\sqrt{\mu}) 
                + \sqrt{\frac{\mu}{\nu}} 
                  {\rm sh}(a\sqrt{\mu}) \rm{ch}[ \sqrt{\nu}(\frac{L}{2}-a) ]
              }
             + \frac{1}{D\mu} 
\end{eqnarray}
and
\begin{eqnarray}
 & & Y_2(x,s) = \frac{1}{D}
                \left( \frac{1}{\mu}-\frac{1}{\nu} \right) \times \nonumber \\
 & & \ \ \ \times   \frac{
                {\rm ch}[(\frac{L}{2}-x)\sqrt{\nu}]
              }{
                {\rm ch}[(\frac{L}{2}-a)\sqrt{\nu}] 
                + \sqrt{\frac{\nu}{\mu}} 
                  {\rm sh}[(\frac{L}{2}-a)\sqrt{\nu}] {\rm ch}(a\sqrt{\mu})
              } + \nonumber \\
 & & \ \ \ \  \ \         + \frac{1}{D\nu} 
\end{eqnarray}
where $Y(x,s)=Y_1(x,s)$ for $0\le x \le a$ and $Y(x,s)=Y_2(x,s)$ for
$a\le x\le L/2$ with $\mu=(s+\sigma_0)/D$ and $\nu=s/D$. The reaction
rate $k(s)$ is given by $k(s)=2 \sigma_0 \int_0^a dx Y_1(x,s)$ and
equals
\begin{eqnarray}
  & k(s) & = \frac{2\sigma_0}{D}
             \left( \frac{1}{\nu}-\frac{1}{\mu} \right) 
             \frac{{\rm sh}(a\sqrt{\mu})}{\sqrt{\mu}} \times \nonumber \\
  & & \times  \frac{
                {\rm ch}(x\sqrt{\mu})
              }{
                {\rm ch}(a\sqrt{\mu}) 
                + \sqrt{\frac{\mu}{\nu}} 
                  {\rm sh}(a\sqrt{\mu}) {\rm ch}[ \sqrt{\nu}(\frac{L}{2}-a) ]
              }
             + \frac{2\sigma_0}{D\mu} 
\end{eqnarray}
We could not find the inverse Laplace transform of the expression above in
closed analytic form. However, this is possible when
$\sigma_0\rightarrow\infty$. In such a case one has
\begin{equation}
  k(s) \approx 2 \sqrt{\frac{D}{s}} \frac{1}{{\rm cth}[(L/2-a)
                       \sqrt{\frac{s}{D}}]} +
                 \frac{2\sigma_0a}{\sigma_0+s} + {\cal O}(1/\sigma_0)
  \label{ks}
\end{equation}

The inverse Laplace transform of the approximate expression for $k(s)$ can
be found by a residuum method. The $s=0$ is not a branching point nor
pole. The only poles come from $cosh$ term in denominator which has
poles at $s_m=-\pi^2(m-1/2)^2D/(L/2-a)^2$. This fully fixes form of
$k(t)$ in Eq.~(\ref{kt}).

\section{Computer Experiment via Monte Carlo Simulations}
\label{app:MCS}

We have chosen the minimal process algorithm for the simulations for
two reasons. The first reason is that the algorithm reproduces the
master equation (\ref{ME}).~\cite{HB} Second, our goal is to study a
whole range of particle sizes and relatively large numbers of
particles at the same time. Clearly, there are another possibilities
to carry out Monte Carlo simulation, but the main advantage of the
minimal process algorithm is that it can be applied for systems
containing relatively large number of particles. An original algorithm
was devised for the situation where $a\sim h$, i.e. particles react at
the same lattice site or when nearest neighbors. We had to modify the
original version of the algorithm to account for finite reaction range
when $a\gg h$. A detailed description of the algorithm is given
bellow.

{\noindent\bf Algorithm:}
\begin{itemize}
\item[(1)] Site $i$ is chosen at random.
\item[(2)] If the site is empty go to step (5).
\item[(3)] For a chosen site $i$, one has to calculate the rate $W_i$
for a certain process to occur (diffusion or reaction). Also, one
needs a null rate $N_i$ where nothing happens (the so called ``null
process''). The null rate is defined from $W_i+N_i=Q$, where $Q$ is
arbitrary but known at each simulation step.  $Q$ is chosen in such a
way that none of the $N_i$ is negative. In practise, the case when $Q$
is taken as the largest of $W_i$ works best since this leads to the
smallest possible values for $N_i$, {\em i.e.} chance that nothing is
done in course of simulation is reduced. (Please note that this
requires that $Q$ is updated as simulation proceeds, but can be done
in a straight forward manner as explained in ref.~\onlinecite{HB})

$W_i=D_i+R_i$ accounts for possibilities that a particle at the site
diffuses to the neighboring site with rate $D_i$, or reacts with a
particle in some other site with rate $R_i= \sum_{j\in\Omega_i}
\sigma(r_{ij})$. $\Omega_i$ denotes set of sites which are within
reaction range of the site $i$. The calculation of $R_i$ is by far the
most costly step when $a$ is large. In that case, a large region has
to be searched in order to find all particles within $\Omega_i$.  This
step costs $M_{search}\sim (a/h)^d$ computational steps if the sites
are checked one by one. The cost can be reduced further by introducing
a list which specifies which sites that contain particles within the
reaction range of the particle at site $i$.  In that case one has to
update the list for each diffusion step made.  The best algorithm we
have so far updates the list in roughly $M_{serach}\sim (a/h)^{d-1}$
steps.
\item[(4)] Once the rates for the site $i$ have been calculated one
can use them to evaluate probabilities for specific process
$p^{(D)}=D_i/Q$, $p^{(R)}=R_i/Q$ and $p^{(null)}=N/Q$.  Once the
probabilities are calculated a certain process is chosen by linear
selection algorithm. First one decides if diffusion, reaction or
nothing is going to happen. If diffusion is to happen then the particle is
moved to one of the randomly chosen $2\times d$ nearest neighbors. If
reaction was chosen, then one of the sites containing particles in
reaction range is chosen at random, {\em e.g.} at site $j$, and pair
of particles from site $i$ and $j$ are annihilated.
\item[(5)] Time is updated according to the formula $t\rightarrow
t+\Delta t$ where $\Delta t=1/L Q$ where $L$ was specified before and
$Q$ is the maximum rate at the present step.
\item[(6)] Move back to (1) unless some criteria to stop is invoked.
\end{itemize}

Applying the same type of reasoning as in ref.~\onlinecite{HB} one can see
that the algorithm proposed here reproduces the behavior described by
the master equation (\ref{ME}).  As time of the simulation progresses
we monitor the number of particles and calculate all the statistics.

As the original minimal process algorithm, the present simulation
method is not that efficient at the later stages of dynamics when the
lattice becomes sparse.  The quantity that governs computational cost
of this method is the number of Monte Carlo steps needed to see some
change in the number of particles. We describe it by the number of
Monte Carlo steps needed to annihilate the {\em last} pair of
particles.

To make such estimate, it is best to move to the reference frame of
one of these particles. Then one particle is fixed and another one is
trying to find it. The number of diffusion steps that the moving
particle needs to find the one who sits still is roughly given by
$M({\rm diff})\sim L^d/a$ (here and in the following it is implicitly
assumed that every length variable is measured in units of lattice
spacing $h$). Each diffusion step bears $M({\rm step/diff})$
computational steps which gives the total number of steps to
annihilate the pair of particles equal to $M({\rm tot})=M({\rm
diff})M({\rm step/diff})$.  The number of Monte Carlo steps per one
diffusion step is roughly 1, $M({\rm step/diff})\sim 1$. However,
calculation of $R_i$ requires updating the internal list which costs
$M_{search}\sim a^{d-1}$ search steps whenever the particle is moved.
Thus, the true number of computational steps per diffusion step is
given by $M({\rm step/diff})\sim M_{search} \sim a^{d-1}$.  Finally,
one gets an estimate for the number of computational steps needed to
annihilate the last pair of particles as $M({\rm tot})\sim L^d
a^{d-2}$.

The algorithm has an interesting property that for $d=1$ there is a
reduction in the computational cost when comparing large and small $a$
cases. For larger $a$ the algorithm works more efficiently. For $d=2$
the computational cost does not depend on $a$.  Simulating a large $a$
situation for $d=3$ is more costly. One could avoid this growing cost
problem at $d=3$ by browsing through particles instead of searching
for sites when calculating $R_i$. This is clearly the preferred option
when the number of particles in the system is not that large.




\end{multicols}       
       

\begin{thebibliography}{20} 

\bibitem{IntRevCyto} Luby-Phelps, K. {\em Int. Rev. Cytol.} {\bf 2000}, 
192, 189.

\bibitem{PhysRev1} Kotomin E.; Kuzovkov, V. {\em Rep. Prog. Phys.}
{\bf 1992}, 55, 2079.

\bibitem{PhysRev2} Kotomin E.; Kuzovkov V. in {\em Comprehensive
Chemical Kinetics}, vol. 34, ``Modern aspects of diffusion-controlled
reactions'', R.G.Compton and G. Hancock Editors, (Elsevier, 1996)

\bibitem{kotkuz} Kuzovkov V.N.; Kotomin, E.A. {\em Chem. Phys.} {\bf 1983}, 
81, 335.

\bibitem{kotkuz1} Kotomin, E.; Kuzovkov, V.; Frank, W.; Seeger, A.
{\em J. Phys. A} {\bf 1994}, 27, 1453.

\bibitem{Chiu1} Chiu, D.T.; Wilson, C.;  Rytts\'en, F.; Str\"omberg, A.;
Karlsson, A.; Nordholm, S.; Hsiao, A.; Gaggar, A.; Garzia-L\'opez, R.;
Moscho, A.; Orwar, O.; Zare, R.N. {\em Science} {\bf 1999}, 283, 1892.
      
\bibitem{Chiu2} Chiu, D.T.; Wilson, C.; Karlsson, A.; Danielsson, A.;
Lundqvist, A.; Str\"omberg, A.; Rytts\'en, F.; Davidson, M.; Nordholm,
S.; Orwar, O.; Zare, R.N. {\em Chemical Physics} {\bf 1999}, 247, 133.

\bibitem{Karl1} Karlsson, A.; Karlsson, M.; Karlsson, R.; Cans, A-S.;
Str\"omberg, A.; Rytts\'en, F.; Orwar, O. {\em Nature} {\bf 2001}, 409, 150.

\bibitem{Karl3} Karlsson, R.; Karlsson, A.; Karlsson, M.; Cans, A-S.;
Voinova, M.; Bergenholtz, J.; Ewing, A.G.; \AA kerman, B.; Orwar, O.
{\em Langmuir} {\bf 2002}, 18, 4186.

\bibitem{Park1} Park, H.; Park, J.; Lim, A.K.L; Anderson, E.H.;
Alivisatos, A.P.; McEuen, P.L. {\em Nature} {\bf 2000}, 407, 57.

\bibitem{spinchains} Alcaraz, F.C.; Droz, M.; Henkel, M.; Rittenberg, V. 
{\em Annals of Physics} {\bf 1994}, 230, 250-302.

\bibitem{KJ1} Konkoli, Z.; Johannesson, H.; Lee B.P. {\em Phys. Rev. E}
{\bf 1999}, 59, R3787.

\bibitem{KJ2} Konkoli, Z.; Johannesson, H. {\em Phys. Rev. E} {\bf 2000}, 
62, 3276.

\bibitem{ChemRev1} {\em Comprehensive Chemical Kinetics}, Vol. 25,
``Diffusion-limited reactions'', C.H. Bamford, C.F.H. Tipper and
R.G. Compton Editors, (Elsevier, 1985).

\bibitem{Khai1} Khairutdinov, R.F.; Serpone, N. {\em Prog. React. Kinetics}
{\bf 1996}, 21, 1-68.

\bibitem{Khai2} Khairutdinov, R.F.; Burshtein, K.Ya.; 
Serpone, N. {\em J. Photochemistry and Photobiology A} 
{\bf 1996}, 98, 1.

\bibitem{Tach1}  Tachiya, M. {\em Chem. Phys. Lett.} {\bf 1980}, 
69, 605.  

\bibitem{Rama} {\em Photochemistry in Organized and Constrained
Media}, edited by Ramamurthy, V.; VCH Publishers (1991).

\bibitem{ben1} Lee, B.P. {\em J. Phys. A} {\bf 1994}, 27, 2633.

\bibitem{ben2} Lee, B.P.; Cardy, J. {\em Stat. Phys.} {\bf 1995}, 80,
971.

\bibitem{fractal} Havlin, S.; Ben-Avraham, D. {\em Adv. in Physics} 
{\bf 1987}, 36, 695.

\bibitem{burl1} Burlatskii, S.F.; Ovchinnikov, A.A.
{\em Russ. J. Phys. Chem.} {\bf 1978}, 52, 1635.

\bibitem{ovch} Ovchinnikov, A.A.; Zeldovich, Ya.B. {\em Chem. Phys.} 
{\bf 1978}, 28, 215.

\bibitem{thwil} Toussaint, D.; Wilczek, F. {\em J. Chem. Phys.} 
{\bf 1983}, 78, 2642.

\bibitem{burl2} Burlatskii, S.F.; Ovchinnikov, A.A.; Pronin, K.A.
{\em JETP} {\bf 1987}, 92, 625.

\bibitem{gut} Gutin, A.M.; Mikhailov, A.S.; Yashin, V.V. {\em JETP} 
{\bf 1987}, 92, 941.

\bibitem{bram} Bramson, M.; Lebowitz, J.L. {\em Phys. Rev. Lett.} 
{\bf 1988}, 61, 2397.

\bibitem{oerd} Oerding, K. {\em J. Phys. A} {\bf 1996}, 29, 7051.

\bibitem{mattis} Mattis, D.C.; Glasser, M.L. {\em Rev. Mod. Phys.} 
{\bf 1998}, 70, 979.

\bibitem{Kopelman} Koo, Y.E.L.; Kopelman, R. {\em J. Stat. Phys.} {\bf 1991},
65, 893-918 

\bibitem{ChemRev2} Calef, D.F.; Deutch, M. {\em Ann. Rev. Phys. Chem.} 
{\bf 1983}, 34, 493.

\bibitem{BiolRev} Berg, H.C.; Purcell, E.M. {\em Biophys. J.} 
{\bf 1977}, 20, 193.


\bibitem{steric} Wu, Y-Ta.; Nitsche, J.M. {\em Chemical Engineering Science}
{\bf 1995}, 50, 1467-1487. 

\bibitem{HB} Hanusse, P.; Blanche, A. {\em J. Chem. Phys.} {\bf 1981}, 
74, 6148.

\end{thebibliography}
\end{document}